# Scalable High-Precision Microfabrication on Various Lithography-Incompatible Substrates and Materials Enabled by Wafer-Scale Transfer Lithography of Commercial Photoresists

Qinhua Guo[1], Zhiqing Xu[1], Lizhou Yang[1], Jingyang Zhang[1], Yawen Gan[1], Jiajun Zhang[1], Jiahao Jiang[1] and Yunda Wang[1, 2, *]

[1] Smart Manufacturing Thrust, The Hong Kong University of Science and Technology (Guangzhou), Guangzhou, China

[2] Department of Mechanical and Aerospace Engineering, The Hong Kong University of Science and Technology, Hong Kong SAR, China

*Corresponding author. E-mail: ydwang@ust.hk

## Abstract

Photolithography conventionally requires flat, rigid and stable substrates, limiting its applications in flexible, curved, and transient electronics. In this study, a breakthrough approach is reported that employs a reversibly adhesion-switchable phase-changing polymer to universally transfer commercial photoresists onto previously inaccessible substrates, overcoming fundamental limitations of conventional photolithography.

## 1 Introduction

Photolithography is an essential technique for fabricating micro- and nanoscale structures, forming the foundation of integrated circuits and MEMS devices [1-4]. However, conventional photolithography requires substrates to be flat and rigid, as light diffraction caused by gaps or deformation of photoresist/substrates affects the pattern fidelity [5]. Moreover, the photoresist coating and development steps involve solvents that can damage solvent-sensitive substrates, causing swelling or degradation [6-8]. These limitations hinder high-precision patterning on unconventional substrates, such as curved or flexible surfaces, three-dimensional microtextured topographies, and delicate material layers including colloidal quantum dot films, where direct spin-coating and photoresist processing are often infeasible [7, 9-12]. Transfer printing of prefabricated functional inks is a revolutionary method to achieve applications of microscale materials and devices in various unconventional contexts, including micro/nano material assembly [13-14], flexible electronics [15-16], curved electronics [17-18], optoelectronics and heterogeneous integration [19-22]. However, these ex-situ fabrication methods also face challenges in general applicability and transfer integrity. The strain and stress of pre-defined functional materials during transfer processes may lead to damage in delicate materials and devices, making serious performance degradation. Additionally, these ex-situ fabrication methods can't meet demands for in-situ microfabrication of photolithography-incompatible materials. Innovative approaches are therefore needed to extend lithographically defined high-resolution in-situ microfabrication to these challenging contexts.

Photoresist transfer printing has emerged as a promising strategy for patterning unconventional substrates. Several approaches have been developed, including detachment lithography [23], PDMS-based kinetic transfer [24], and tape-assisted transfer [5, 25-27], each with distinct mechanisms and associated limitations. **Detachment lithography** relies on mechanically induced fracture of a continuous photoresist film to create designed patterns [23]. In this method, a PDMS stamp coated with photoresist film is brought into contact with a structured silicon mold, and then rapidly retracted to make the photoresist film broken along protruded features, forming patterned photoresist structures on the PDMS stamp. The patterned photoresist can subsequently be released onto target receiver substrates following conformal contact and low-speed separation processes. While this method allows patterning on both planar and curved substrates, it requires pre-fabricated molds with customized topographies, and the sudden mechanical stress often results in uncontrolled fracture, limiting its yield and scalability. **PDMS carrier-based methods** rely primarily on peeling dynamics to kinetically modulate adhesion of elastomer stamp, facilitating photoresist transfer from low-surface-energy donor substrates to unconventional receiver substrates [24]. However, this kinetic control can result in risks of deformation-induced fracture and limited adhesion contrast [28-29]. **Tape-assisted transfer** employs thermal release tapes to transfer custom-formulated photoresist, often incorporating surfactants to modulate photoresist adhesion to the donor substrate [5, 25-26]. This method enables dry patterning on unconventional substrates. However, the reliance on customized photoresists may limit generality and scalability, making the approach less suitable for broader adoption. Moreover, thermal release tapes are generally single-use due to their irreversible adhesion transition, which limits their reusability in scalable fabrication. Their limited conformability can also pose challenges for patterning on textured or non-planar surfaces. Additionally, transfer accuracy and wafer-scale registration have not been systematically demonstrated in these methods, leaving open challenges for precision-critical applications.

## 2 Results and Discussion

### 2.1 Working principle of the SPRR polymer-based photoresist transfer method

Figure 1 shows the working principle of the wafer-scale photoresist transfer method developed for patterning unconventional substrates. This method employs a sharp phase-changing rigid-to-rubbery polymer (SPRR polymer) to transfer commercial photoresists from low-surface-energy-treated donor substrates (e.g., PDMS-coated surfaces) to various unconventional substrates [30], including those that are stretchable, flexible, curved, or otherwise susceptible. As shown in Fig. 1a, this design enables the huge storage modulus change range of the SPRR polymer over a narrow phase change temperature range, exhibiting a large, reversible change. Figure 1b demonstrates the rigid-to-rubbery transition of SPRR polymer heated by a hotplate. The changes in storage modulus fundamentally determine the adhesion-switching behavior of SPRR polymer. This thermal responsiveness, accompanied with a shape memory polymer (SMP)-like behavior, enables controlled transitions between a soft, conformable state and a rigid, dimensionally stable state during the pickup and release steps. The low-surface-energy PDMS coating on the rigid donor substrate is applied via a scalable spin-coating process, while the transferable photoresist is prepared using standard photolithography.

### 2.2 High-precision, wafer-scale photoresist transfer

To evaluate the effectiveness and scalability of our method for patterning unconventional surface at the wafer scale, we conducted an experiment involving the transfer of photoresist from a 4-inch PDMS-coated wafer onto a 4-inch solvent-susceptible substrate which is PVA. As shown in Fig. 2a, multiscale patterns (5–50 µm features) of 1-µm-thick SU-8 2002 photoresist were photolithographically defined on a 4-inch PDMS-coated silicon wafer (PDMS thickness: 15 µm). A 4-inch SPRR polymer/PET carrier (350-µm-thick SPRR polymer coating on 100-µm-thick PET) was laminated onto the donor wafer using a commercial hot laminator. During the lamination process, the flexible SPRR polymer made conformal contact with the 4-inch donor wafer, while the smooth PET film minimized a shear force from the roller. After cooling to room temperature, the photoresist was picked up by separating the carrier at an average fracture propagation speed less than 2 mm $s^{-1}$. For photoresist release, the SPRR polymer/PET carrier with photoresist was laminated onto a 4-inch, 100-µm-thick PVA film and peeled away on an 80 °C hot plate at the average fracture propagation speed less than 2 mm $s^{-1}$, completing the transfer.

Figure 2c shows an overlaid processed image of the discrete photoresist structures before and after the wafer-scale transfer. The alignment was achieved using a linear transform with assistance of Fiji software [33-34]. The local registration error, measured as deviations in relative position of photoresist structures, was within 1.8±0.9 µm for translation and below 0.03±0.03 radians for rotation across a 1380 µm × 650 µm area. To our knowledge, these registration results represent the highest accuracy reported to date for photoresist transfer. This is attributed to the high modulus of the SPRR carrier in its rigid state, which effectively locks the photoresist structures before release, and its soft state, which enables damage-free release.

### 2.3 Transfer of commercial photoresists onto diverse unconventional substrates

To demonstrate the general applicability of our method, we conduct a series of experiments transferring various commercial photoresists onto different receiver substrates. The transfer protocol followed the same framework, comprising pickup and release steps. The donor substrates were prepared by patterning commercial photoresists via photolithography on PDMS-coated wafers. In these experiments, the SPRR polymer/glass carrier with a 1-mm-thick SPRR layer was used, and the heating step was performed on an 80 °C hot plate. The average fracture propagation speeds during pickup and release were estimated to be below 2 mm $s^{-1}$.

Figure 3a–h show 3-µm-thick discrete AZ5214E photoresist (AZ PR) structures with 10 µm feature sizes transferred onto a variety of unconventional substrates. Figure 3a–d illustrate successful photoresist transfers onto several flexible films, including polyimide (PI), polyethylene terephthalate (PET), silicone gel film, and polyurethane (PU). Additionally, the method is compatible with substrates that are incompatible with conventional solution-based lithographic processes. Figure 3e and f show photoresist transfers onto a water-soluble PVA film and a fluoropolymer film with a hydrophobic surface, respectively. Figure 3g and h show brightfield and fluorescence images, respectively, of AZ5214E photoresist transferred onto a $CsPbBr_3$ quantum dots/glass substrate.

The method also supports transfer of thick photoresists. Figure 3i shows the SEM image of discrete 10-µm-thick SU-8 2010 photoresist structures transferred onto a free-standing silicone gel-film. In addition to discrete features, the method enables transfer of continuous photoresist films. Figure 3j shows the SEM image of 10-µm-thick continuous film of patterned SU-8 2010 photoresist transferred onto a free-standing PVA substrate. Extending photolithography to curved surfaces has been a long-standing challenge due to the incompatibility of rigid masks and spin-coatings with non-flat geometries [35-36]. Using the SPRR polymer-based transfer method, we successfully demonstrated high-resolution photoresist patterning on curved convex surfaces. Figure 3k and l show discrete structures and a continuous film of SU-8 2010 photoresist, respectively, transferred onto curved substrates using a 1-mm-thick free-standing SPRR polymer carrier. Further details on material preparation and photoresist transfer processes are provided in the Experimental Section.

## 3 Conclusions

We introduce a high-fidelity and scalable photoresist transfer method employing a sharp phase-changing rigid-to-rubbery polymer carrier that has a nearly 2300-fold change in storage modulus across a moderate melting temperature. During rigid-to-rubbery transition, the phase-changing polymer exists significantly adhesion-switchable capability enabling the transfer of photoresist from PDMS-coated donor wafer onto previously inaccessible substrates. Utilizing controlled transition between rigid, dimensionally stable state and soft, conformable state during alignment, pickup and release steps, this transfer method enables a wafer-scale (~4-inch) photoresist transfer with global registration error below 60 µm.

Unlike previously reported photoresist transfer method, our approach demonstrates general applicability for multiple commercial photoresists that are ready-to-use and not reliant on additional treatment. These photoresists, defined through standard photolithography, were successfully transferred onto a broad range of substrates previously incompatible with lithography, including flexible film, solvent-sensitive layers, curved and microtextured surfaces, and delicate materials. These transfer experiments highlight the generality and scalability of our method for broader adoption.

This study systematically demonstrates the application potential of the photoresist transfer in various scenarios that are inaccessible to conventional lithography, opening new opportunities for groundbreaking applications of high-resolution microfabrication in emerging fields, such as flexible electronics, paper-based electronics, curved electronics, transient electronics, and optoelectronics.

## Acknowledgments

The authors want to thank Prof. Qingming Chen at Sun Yat-sen University and Prof. Shengdong Zhang at Peking University for their valuable discussion. Also, the authors want to thank the support in equipment from the following lab in The Hong Kong University of Science and Technology (Guangzhou): Wave Functional Metamaterial Research Facility (WFMRF), Materials Characterization and Preparation Facility (MCPF), Advanced Additive Manufacturing Laboratory (AAM), The Center for Heterogeneous Integration of μ-systems and Packaging (CHIP), and Laboratory for Brilliant Energy Science and Technology (BEST Lab). The authors also want to acknowledge the equipment support from the laboratory of School of Microelectronics Science and Technology in Sun Yat-sen University. ChatGPT was used to refine the English in the manuscript.

## Declarations

The manuscript does not contain clinical studies or patient data.

**Conflict of Interest** Authors declare that they have no competing interests.

**Data Availability Statement** All data are available in the main text or the supplementary materials.